# Transforming common III-V and II-VI semiconductor compounds into topological heterostructures: The case of CdTe/InSb superlattices

Qihang Liu[*], Xiuwen Zhang, L. B. Abdalla and Alex Zunger[*]

University of Colorado, Boulder, Colorado 80309, USA

[*]E-mail: qihang.liu85@gmail.com, alex.zunger@colorado.edu

**Abstract**

Currently known topological insulators (TIs) are limited to narrow gap compounds incorporating heavy elements, thus severely limiting the material pool available for such applications. We show via first-principle calculations how a heterovalent superlattice made of common semiconductor building blocks can transform its non-TI components into a *topological nanostructure*, illustrated by III-V/II-VI superlattice InSb/CdTe. The heterovalent nature of such interfaces sets up, in the absence of interfacial atomic exchange, a natural internal electric field that along with the quantum confinement leads to band inversion, transforming these semiconductors into a topological phase while also forming a giant Rashba spin splitting. We reveal the relationship between the interfacial stability and the topological transition, finding a "window of opportunity" where both conditions can be optimized. Once a critical InSb layer thickness above ~ 1.5 *nm* is reached, both [111] and [100] superlattices have a relative energy of 5-14 meV/A$^2$ higher than that of the atomically exchanged interface and an excitation gap up to ~150 meV, affording room-temperature quantum spin Hall effect in semiconductor superlattices. The understanding gained from this study could broaden the current, rather restricted repertoire of functionalities available from individual compounds by creating next-generation super-structured functional materials.

## 1. Introduction

Topological insulators (TIs) are nonmetallic 3D bulk compounds or 2D nanostructures having an inversion in the order of the valence and conduction bands at special, time reversal invariant wave vectors in the Brillouin zone. While band inversion *per se* is not new, and has been recognized long ago in common semiconductors and semi metals, such as HgTe[1], $\alpha$-Sn[2] or PbTe[3], what is new is that this effect leads in lower dimensional forms of the parent 3D or 2D structures to 2D surfaces states and 1D edge states, respectively, that possess passivation-resistant, linearly dispersed metallic energy bands [4], and to the quantum spin-Hall effect (QSHE) consisting of counter-propagation of opposite spins in spatially distinct channels in the absence of a magnetic field. TIs have recently attracted interest due to their ability to realize novel quantum phases including such quantum spin Hall effect [5, 6], quantum anomalous Hall effect [7], and topological superconductivity [8], and due to their promise of future potential applications in spintronics [9] and quantum information [10].

The required band inversion[6, 11] in the primary 3D or 2D systems is generally achieved by introducing high atomic number (Z) heavy elements, manifesting strong spin-orbit coupling (SOC) [12]. The required incorporation of high-Z elements such as Hg, Tl, Pb, and Bi limits the available material pool relevant for discovery of such TI's and often makes it difficult to achieve thermodynamically stable structures that are, at the same time topological. In the short history of TI search, many theoretical predictions of TI compounds in postulated, hypothetical structures did not correspond to the thermodynamically stable, or near-stable forms of these compositions[13]. Especially, after the prediction of QSHE in graphene[5], numerous proposals[14, 15] borrowed the 2D layered structure with assumed composition by heavier atoms without examining the structure with respect to alternative phase, yet optimistically reporting new TI functionality and large excitation gap. For example, the assumed 2D structures of BiBr was predicted[15] to have an impressively large excitation gap ~ 0.74 eV, but corresponds according to the present first-principles calculations to a high energy phase of this compound (the ground state 3D structure of this compound listed in ICSD[16] is 150 meV/atom lower in energy and has an excitation gap of only 0.025 eV). Similarly, many

*III*BiO$_3$ compounds (*III*= Al, Ga, Y..) that were predicted to be TI turn out to be so only in a high energy cubic structure, far in energy above the non-cubic stable structures that are not TIs[13]. Whereas stabilization of high energy structures is certainly possible sometimes, this generally requires special procedures and the resulting structure may not be ideally suited for robust device applications or room temperature. It does make sense therefore to engage in co-evaluation of the target functionality and the structural stability of the phase said to have the desired functionality.

Such difficulties in realizing topological structures of individual building block compounds have encouraged to exploration of approaches to converting non-TI compounds into TI compounds by manipulating electronically or mechanically the band structure of a *single compound* (e.g by external electric field [17] or by applied external strain[18], and more recently by designing a *combination of different material* building blocks into nanoscale superlattices or quantum wells, utilizing *heterostructure effects* such as built-in electric field and quantum confinement[6, 19-21]. The advantage of using heterostructures that involve conventional II-VI, III-V or group-IV semiconductors is that they come with well-studied growth methodologies and can be readily integrated in current device technology. The prototypical examples of proposed heterostructure TI include (i) Combination of an already band-inverted bulk semimetal acting as quantum well (e.g., HgTe) with a normal bulk insulator acting as barrier (e.g., CdTe), as in the isovalent II-VI/II-VI quantum heterostructure HgTe/CdTe [6, 22]. Here the transformation occurs from a non-topological system with a thin quantum well to a topological heterostructure once the well thickness exceeds a critical value [22, 23]; (ii) Combination of two *isovalent* and nonpolar normal insulators such as InAs and GaSb[19] manifesting a type-III "staggered" band alignment [24] in which the conduction band minimum of one component (InAs) lies below the valence band maximum of the second component (GaSb). Here reduction in the InAs well thickness creates band inversion, (however with band *anti*-crossing [25]); and (iii) Combination of two *polar* (wurtzite) insulators (such as InN/GaN[26]) with polar interfaces that can produce built-in electric fields, leading to an field-induced band inversion. However, isovalent heterostructures such as InN/GaN where the built-in electric field is from the spontaneous and piezoelectric polarization effect in wurtzite structure[26] requires a significant lattice mismatch (such as 10% in InN/GaN) to set up the required electric field. In the case of InN/GaN a period of at least 4 monolayers (about 1 *nm*) was suggested theoretically, a value that is well above the threshold thickness (about 0.6 *nm* for 10% mismatch)[27] for nucleating strain-induced dislocations, making it thermodynamically difficult to grow. Regarding artificially grown 2D structures, to the best of our knowledge, QSHE was observed thus far only in CdTe/HgTe[28] and InAs/GaSb[29] quantum wells [prototype (i) and (ii)] by means of low-temperature electron transport.

In the present paper we propose to create topological structures from non-topological components via 2D layering that create a built-in electric field due to charge mismatch between the heterovalent components at interface, instead of the field supplied by the polar nature of bulk components as in InN/GaN[26]. In such *heterovalent superlattices* one combines two normal insulators that belong to different valence classes creating a *heterovalent* heterostructure. The first example of such heterovalent system is the Ge/GaAs quantum well proposed by Zhang *et al.* that has been theoretically predicted to be topological due to heterovalent electric field in the zincblende structure[20]. Here we bring in a new prototype of heterovalent zincblende compounds that can transform from semiconductors to topological insulators: The III-V/II-VI class. This class offers enormous freedom to form different heterovalent alloys, monolithically integrated planar heterostructures, and quantum-dot structures, and thus presents novel physical properties, different from those of the isovalent heterostructures. Here, the stability issue takes up a special form: atomic exchange across the interface in heterovalent interfaces is generally an energy-stabilizing event and if carried to completion will cancel the electric field and improve the thermodynamic stability[30-33]. Whether such atomic exchange can be complete (resulting in a structure incapable of converting the system to a TI) or only partial (resulting in a structure capable of converting the system to a TI albeit with higher energy) is often an open question, as actually grown samples [34-36] can be either. Based on first-principle calculations, we use $(InSb)_m/(CdTe)_n$ heterostructure as an illustration to establish the relationship between the interfacial structure, stability and the topological transition in heterostructures made of common-semiconductor building blocks. Such III-V/II-VI heterostructures have been grown successfully by molecular beam epitaxy (MBE)[36, 37] at an optimized temperature 310 ºC, but were not considered as candidates for TI.

We find that without interfacial atomic exchange such heterovalent superlattices allow field-induced transformation into a band inverted topological phase above critical InSb layer thickness *m*. For [111] structures this transition requires a well thickness above ~ 4 monolayers (4 ML, or 1.5 *nm*) for a thick CdTe barrier, an excitation band gap 8 meV in TI phase. In addition, giant Rashba spin splittings with helical spin textures are manifested in the valence subbands. Thermodynamically, such topology-inducing superlattices with abrupt interfaces have energies ~ 5 meV/A$^2$ or more above the atomically exchanged ground state -- a moderate degree of metastability that might be achieved experimentally. On the other hand, [100] superlattices without interfacial atomic exchange are higher in energy than [111] by ~ 8 meV/A$^2$, and also become TI above a critical InSb thickness *m* ~ 4 with an extended excitation band gap up to 156 meV, which is desirable for realizing quantum spin Hall effect at room temperature. This work illustrates how to make realistic predictions on TI by co-evaluating the competition between thermodynamic stability and band inversion, not just aiming at realizing a target property in assumed hypothetical structures that could be thermodynamically unrealizeable. The theoretical discovery of TI-ness among ordinary binary octet semiconductors III-V and II-VI opens this technologically well-studied group to the realm of topology.

## 2. Basic interfacial physics of III-V/II-VI heterovalent superlattices
### 2.1. Changes in thermodynamic stability due to interfacial atomic exchange

In heterovalent superlattices the total energy consists of three terms: (i) The energy increase induced by the formation interfacial "wrong bonds" $2N(G)\delta$, such as super-octet (nine electron III-VI) bonds and sub-octet (seven electron II-V) bonds. Here *N(G)* denotes the number of wrong bonds in configuration *G* (growth orientation and reconstruction pattern) and $\delta$ denotes the average bond energy of the two bonds types; (ii) The increase in electrostatic energy due to charge interfaces induced by the electrostatic Madelung potential $q\Delta E_C(G)$, proportional to the excess charges *q* at the interfaces; and (iii) The strain energy [38] $\Delta E_S(G)$ due to the tensile or compressive strain of the possibly lattice-mismatched constituents. A model considering the energy of these three terms fit to density functional theory (DFT) ingredients was developed by Dandrea *et al.* and applied to IV-IV/III-V superlattices [31]:

$$\Delta H = 2N(G)\delta + q\Delta E_C(G) + \Delta E_S(G). \tag{1}$$

Once fit to DFT results, for each layer orientation [hkl] one can search for different interfacial atomic patterns (reconstruction) that minimize the energy. For example, within a 2x2 supercell *the abrupt* [111] interface has 4 wrong bonds along with a strong electrostatic field; whereas a fully atom exchanged [111] interface has 6 wrong bonds but vanishing excess charges (see Figure 1a and b). Clearly, atomic exchange sets up a competition between the different terms in Equation 1. Such understanding clarifies the roles of these different factors and allows a quick estimation of formation energies of various complex configurations.

**2.2. Changes in energy bands due to the internal electric field**

Without atomic exchange, *abrupt* III-V/II-VI superlattices have a charge imbalance ("polarity") at both interfaces: One interface has *super-octet III-VI bonds* (*donors,* with +1/4 electron charge), while the other interface has *sub-octet* II-V *bonds* (*acceptor,* with -1/4 electron charge). This abrupt interface implies a two-dimensional electron gas (2DEG) at one interface and a two-dimensional hole gas (2DHG) at the other. As a result, such abrupt interfaces would have a built-in electric field experienced by both III-V and II-VI layers, resulting in the Stark effect that causes the valence band maximum (VBM) and conduction band minimum (CBM) to bend towards each other. The modification of energy levels by the natural internal electric field is superposed with the quantum confinement effect, in which the energy levels of electrons are shifted up and those of holes are shifted down at reduced layer thickness. Here we will take advantage of this intrinsic polar field modified by quantum confinement to design band inversion and Rashba spin splitting.

Given that the built-in electric field is desirable for affecting transformation from normal to topological insulators but could lead to higher-energy structures, our design objective is to find heterovalent III-V/II-VI superlattices with sublayer thicknesses and orientations that have sufficient electric field to create a NI-TI transition but are thermodynamically not too high in energy.

## 3. Interficial structure and stability of abrupt (polar) and charge-compensated (nonpolar) [111] configurations

We introduce first, in some detail the $(InSb)_m/(CdTe)_n$ superlattices grown along the [111] direction, while the [100] and [110] superlattices will be considered later.

Within a 2x2 supercell, two types of the interfaces are considered. The first configuration has no atomic exchange, being an *abrupt interface*. As shown in Figure 1a, there are four In-Te super-octet wrong bonds in one interface and four sub-octet Cd-Sb wrong bonds at the other (so in Equation 1 we have $N([111]_p) = 4$). These excess charges (see Figure S1 of Supplementary Materials for the charge density plot) would raise the total energy due to their repulsive electrostatic potential. Therefore, such abrupt heterovalent configuration could attempt to reconstruct by swapping group-V atoms with group-VI atoms of the donor interface, thus lowering the formation energy [31-33]. As shown in Figure 1b, after exchange of one Sb atom (out of 4) at one interface with one Te atom at the other, both interfaces now have gone up to three super-octet bonds and three sub-octet bonds, (so energy in Equation 1 is raised by $N([111]_{np}) = 6$**)**, while creating two charge-compensated nonpolar interfaces (so energy is lowered in Equation 1 by $q = 0$). In this case there is no extra charge at any interfaces and thus no built-in electric field. In polar configurations there are built-in electric fields applied on InSb layers, while in nonpolar configurations there are no electric fields but a potential step between InSb and CdTe, as shown in Figure 1c and d. The *nonpolar* configurations were theoretically predicted to be the most stable structure among different interface configurations in certain III-V/II-VI superlattices [33] such as $(GaSb)_6/(ZnTe)_6$. However, the abrupt polar configuration can be lower in energy than the atomically exchanged configuration for a thinner well width (see below).

Next we will focus on the thermodynamic stability of these two configurations *as a function of the layer thicknesses* (*m* and *n*). We note that this atomic exchange could happen at either anion layer (Sb replaced by Te), or cation layer (Cd replaced by In), and each kind of atomic exchange could occur at different in-plane relative positions at the interface, forming totally 8 possibilities within a 2x2 supercell. Both our calculation and previous results [33] give very similar energy (within 0.5 meV/A$^2$) between these configurations, so we use the anion exchange configuration as a representative of nonpolar configurations.

The stability of heterovalent superlattices towards phase separation, i.e., the left-hand side of Equation 1, is given by the formation enthalpy and evaluated here by first-principles total-energy calculation:

$$\Delta H(m, n, G) = 2S\Delta E_{int}(m, n, G) = E_{tot}(A_m B_n, G) - [mE_{tot}(A) + nE_{tot}(B)], \quad (2)$$

where $G$, $S$, $E_{tot}$, $A$ and $B$ stand for the grown direction, interfacial area, total energy, InSb and CdTe, respectively. The interfacial energy $\Delta E_{int}$ (per Å$^2$) is relevant to stability because the energy variation occur mostly at the interface. We focus on [111] (InSb)$_m$/(CdTe)$_n$ superlattices with the thickness of InSb $m$ = 1-6, and fix $m + n$ = 12. The experimental lattice constant of InSb and CdTe is 6.47 and 6.48 Å at 300 K, respectively), indicating nearly perfect lattice match (only 0.15% mismatch) so the third term of the right-hand side of Equation 1 is negligible for our system. Therefore, we fix the lattice parameter of the superlattices as 6.47 Å, using a 2x2x12 supercell, and relax all the internal degrees of freedom inside the cell. Figure 2a shows the interfacial energy $\Delta E_{int}$ as a function of $m$ for both polar and non-polar configurations. Several observations can be made:

(i) The formation energy of either abrupt (polar) or reconstructed (nonpolar) configurations w.r.t. the binary components is *positive*, that is thermodynamically unstable w.r.t. phase separation. This is the standard case for almost all semiconductor superlattices [31, 33, 38]. Nevertheless, such superlattices can still be grown [39] provided the formation energy is not too large. All the interfacial energies are within 3-6 meV/Å$^2$, which is lower than that in ZnTe/GaSb [33]. Indeed, both InSb/CdTe and ZnTe/GaSb heterostructures have been successfully synthesized [34, 36].

(ii) For reconstructed *nonpolar* configurations the formation energy is nearly unchanged with increasing InSb layer thickness. This is because all such thicknesses have full charge compensation at interfaces, so the formation energy is mostly contributed by wrong bonds. However, their number remains unchanged with different InSb layer thicknesses.

(iii) For abrupt polar configurations, both wrong bonds energy and electrostatic energy contribute and, in fact, compete: For *short* InSb well width $m$ = 2, the excess positive and negative charge can easily transfer from III-VI bonds to II-V bonds across the thin InSb well and compensate each other, leading to $q\Delta E_C < 2\delta$ and thus the abrupt configuration is lower in energy than the reconstructed nonpolar configuration. On the other hand, for thicker wells $m > 2$, the excess charge increases with $m$ leading to $q\Delta E_C > 2\delta$, causing abrupt interface to acquire higher energy as $m$ increases. In addition, the formation energy of abrupt interface reaches saturation at large $m$, following the trend of the excess charges (1/4 per bond).

(iv) The abrupt configuration is indeed higher energy than the reconstructed configuation for thicker wells $m > 2$, but the energy difference is moderate: 2.5 meV/A$^2$ for $m = 6$. Considering that in layer-by-layer MBE growth atomic exchange between two interfaces may be an activated process, it is possible that depending on growth temperature and growth rates the abrupt interfaces or the partially compensated interfaces can be stabilized during growth. Such structures will have finite built-in electric field that can be utilized to design band inversion, as discussed next.

## 4. Transforming non-topological compounds to topological structures in abrupt [111] heterovalent superlattices

Due to the absence of the built-in electric field, the reconstructed nonpolar InSb/CdTe superlattices have well-defined band offsets. DFT calculation shows a normal type-I band alignment, and the band offsets vary depending on the type of atom swapping (e.g., Sb-Te or In-Cd swapping, see Figure S2). The InSb layer acts as a quantum well whereas the CdTe layer acts as barrier. Starting with thin InSb well and increasing its thickness $m$ reduces the direct band gap due to the quantum confinement (see Figure 2b). When the InSb well is thick, the band gap approaches the value of bulk InSb.

A system that is not fully charge compensated has an internal electric field. As a result, the band structure is modulated, causing the VBM and CBM to move towards each other and finally inverting their order for thicker InSb layers (increasing $m$). The topological transition is schematically explained by Figure 3a-c. For the interface between InSb well and CdTe barrier, the InSb layer with more super-octet bonds has excess electrons (*n* doped,) while the InSb layer with more sub-octet bonds has excess holes (*p* doped). These layers are denoted as (InSb)$^+$ and (InSb)$^-$, respectively in Figure 1a. Therefore, the InSb well forms effectively a *p-i-n* junction, as shown in Figure 3a.

The built-in electric field applied on the InSb well leads to a potential difference between $(InSb)^+$ and $(InSb)^-$. When the potential difference is large enough, the CBM of $(InSb)^+$ is lower than the VBM of $(InSb)^-$, leading to an overlap between the two bands in *k*-space and eventually to the inverted band order of Γ (denoted in Figure 3b). Such potential difference in InSb layer increases with the ascending layer number *m*, as shown in Figure 1c, and thus leads to the topological phase transition as the thickness exceeds a critical value. We define the *inversion energy* at Γ as $\Delta(\Gamma) = E_6 - E_8$, and thus a system becomes a TI when $\Delta(\Gamma)$ is negative. At the *k*-points off Γ, SOC lowers the band symmetry and thus opens an insulating gap $E_g$. This excitation band gap denoted in Figure 3c is important for realizing the quantum spin Hall or quantum anomalous Hall effect at room temperature.

The band structure of abrupt polar InSb/CdTe superlattice for well width *m* = 3 below the critical thickness for conversion to TI is shown in Figure 4a. We find that as is the case in bulk InSb, there is a direct band gap located at the Γ point with the CBM composed of the $\Gamma_6$ s-like state and the VBM composed of the $\Gamma_8$ p-like state. Furthermore, by projecting the eigenstates onto each atom in the real-space we find that the CBM and the VBM are dominated by the two sides of the InSb well: $(InSb)^+$ and $(InSb)^-$, consistent with our schematic analysis in Figure 3a. This band order is inverted by the weakening quantum confinement and enhanced Stark effect when the well thickness increases to *m* = 5, as shown in Figure 4b, indicating a TI phase. The inversion energy $\Delta(\Gamma)$ as a function of the InSb well thickness for abrupt superlattices is shown in Figure 2b. The critical point for band inversion occurs beyond *m* ~ 4, corresponding to 1.5 *nm* InSb well thickness. To confirm the relationship between the topological nature and band inversion, we further calculate the topological invariant $Z_2$ by tracking the evolution of the Wannier charge centers (WCCs, see Methods for details) in these non-centrosymmetric systems [40, 41]. Given an arbitrary reference line (green line), for *m* = 3 the number of transitions of WCC is even (in this case, 0, see Figure S3a), indicating a normal insulator. In contrast, for *m* = 5 there are odd number of WCC transitions (in this case, 1, see Figure S3b). Therefore, $Z_2$ jumps from 0 to 1, confirming a transition from NI to TI above a critical thickness.

The excitation gap $E_g$ of the InSb/CdTe superlattice for well thickness above topological transition is about 8 meV, corresponding to a temperature limit ~ 90 K for realizing the quantum spin Hall effect. In contrast, for reconstructed nonpolar configuration the inversion energy Δ(Γ) (equals to $E_g$) is always positive because of the lack of built-in electric field (see Figure 2b). The distinct comparison between abrupt and reconstructed configurations suggests that the band inversion found here is induced by the intrinsic electric polarization in polar interfaces, while the excitation gap off the Γ point is caused by the effect of SOC. Such field-induced topological phase transition opens more possibility to create TI using conventional zinc-blende compounds and thus expands the hitherto limited material base of TI.

**5. Giant Rashba spin splitting in sub bands of abrupt superlattice**

The normal zinc-blende semiconductors InSb and CdTe have nonpolar $T_d$ space group, and thus is expected to manifest Dresselhaus splitting [42] rather than Rashba splitting [43] (distinct by the spin textures). However, by taking advantage of the intrinsic polar field in abrupt $(InSb)_m/(CdTe)_n$ one can design and tune the intriguing Rashba splitting [44] in such system. Associated with large SOC and electric field, Rashba effect is connected to many novel phenomena and potential applications, such as spin field effect transistor (SFET) [45], intrinsic spin Hall effect [46], and Majorana Fermions [47]. In the presence of electric field $E$ along $z$ direction, the Rashba-type interaction is described by a momentum-linear Hamiltonian:

$$H_R = \lambda(E \times \boldsymbol{p}) \cdot \boldsymbol{\sigma} = \alpha_R(\sigma_x k_y - \sigma_y k_x), \qquad (3)$$

where $\boldsymbol{p}$ and $\boldsymbol{\sigma}$ denote electron momentum and Pauli matrix vector $(\sigma_x, \sigma_y, \sigma_z)$, respectively. After the inclusion of Equation (3), the typical two-fold spin degenerate band splits into two spin polarized branches (see black frames in Figure 4b), and the wavefunctions of the two branches correspond to electrons with spins oriented in opposite directions perpendicular to the wave vector.

The band structure for all $m$ = 1-6 $(InSb)_m/(CdTe)_n$ [111] abrupt superlattices is shown in Figure S4. We find obvious Rashba-like band splitting in the subbands about 200 – 300 meV below VBM for all the superlattices. The characteristic features quantifying the strength of Rashba effect is Rashba energy $E_R$ defined by the energy difference between the band peak and the crossing point, the corresponding momentum offset $k_R$, and Rashba parameter $\alpha_R$ (defined by $2E_R/k_R$). Figure 5a shows both $E_R$ and $\alpha_R$ of the subbands as a function of $m$. We find that both $E_R$ and $\alpha_R$ increase monotonically as the InSb well becomes thicker, because these two subbands are dominated by two InSb layers at the side of sub-octet bonds (see Figure 4a and b), and more excessive charge can survive against transferring to the side of super-octet bonds.

The magnitude of $\alpha_R$ from $m = 3$ is in the range of 2-4 eV•Å, which is one of the largest values among the Rashba effects currently found in different materials (e.g., 3.8 for BiTeI [48] and 4.2 for GeTe [49]) and at least one order larger than that of the conventional homovalence heterostructures (e.g., InGaAs/InAlAs quantum well) [50]. The spin textures of the two subbands are shown in Figure 5b and c. We found two sets of helical spin propagating oppositely to each other, which is the fingerprint of Rashba splitting. Therefore, the emergence of giant Rashba effect of holes, by moving the Fermi level onto these valence subbands in *p*-doped environment, is expected for spintronic applications.

## 6. InSb/CdTe superlattices along [100] and [110] directions

Having described above the general ideas of the thermodynamic stability vs topological physics of abrupt polar (InSb)$_m$/(CdTe)$_n$ [111] superlattice, we next consider this heterovalent superlattice grown along [100] and [110] direction. Similar with [111] direction, [100] superlattice with an abrupt interface has excess charges at the super-octet and sub-octet interface and thus built-in electric field. This polar field could be fully compensated by atomic exchange. On the other hand, [110] superlattice is already charge compensated and thus expected to be stable against reconstruction, so we will not artificially create the built-in electric field by atomic exchange. Figure 6a exhibits the interfacial energy $\Delta E_{int}$ for different configurations of the three directions for (InSb)$_m$/(CdTe)$_n$ superlattices as a function of $m$. We can categorize the curves into two classes: for polar interfaces, $\Delta E_{int}$ increases with a saturation when $m$ increases; while for nonpolar interfaces $\Delta E_{int}$ remains nearly unchanged, as discussed in Sec. III. The energy order of nonpolar interfaces is [110] > [111] > [100], which is determined by the areal density of wrong bonds, i.e., $2\sqrt{2}/a^2 < 2\sqrt{3}/a^2 < 4/a^2$ (*a* is the in-plane lattice parameter) for [110], [111] and [100] direction, respectively. In polar configurations, [100] interfaces have larger wrong bonds density and excessive charge density, so they are high in energy than [111] polar interfaces.

Figure 6b shows the transition of the inversion energy $\Delta(\Gamma)$ of InSb/CdTe abrupt superlattices with increasing thickness of the InSb layers. For nonpolar [110] configuration, $\Delta(\Gamma)$ is determined only by quantum confinement. Therefore, there is no topological transition in such nonpolar interface because the lower limit of the band gap is the bulk value of InSb. In contrast, [100] abrupt configurations have a NI-TI transition between $m = 3$ and 4 induced by the built-in electric field. When the bands are strongly inverted [a large negative $\Delta(\Gamma)$], multiple band inversions could happen within conduction or valence bands, leading to the upturn in [100] direction seen at $m = 5$. The band structures of [100] abrupt configurations before ($m = 2$) and after ($m = 4$) topological transition are shown in Figure 4c and d, respectively. We find that the excitation gap for $m = 4$ is 156 meV, much larger than that of [111] abrupt TI configurations, while comparable with the value of a recent theoretical proposal of InSb $p$-$i$-$n$ junction (~0.1 eV)[21]. Such a large gap is favorable for realizing quantum spin Hall effect at room temperature. However, comparing to other configurations [100] abrupt configurations are thermodynamically higher in energy (13.6 meV/A$^2$ for $m = 4$). Basically, one can expect partial atomic exchange to get a compromise of stability and TI-ness, i.e., the residual field can convert the system to a TI with a relatively low thermodynamic energy. Such actual samples and interface characterization is called for.

## 7. Discussion and Conclusion

When a bulk grown compound is significantly (say, hundreds of meV/atom) higher in energy than its competing phases (such as decomposition products), there is the possibility that it will not be the phase that actually grows since the competing phases can grow instead. On the other hand, in layer-by-layer growth from the gas phase (as in MBE or MOCVD superlattice growth) the multilayered structure is growable if its energy is above that of competing phases by only small amount (say, less than 100 meV/atom). Furthermore, in layer-by-layer growth, once made, the multilayer structure is rather robust against transformation to other competing phases at room temperature because this often entails the energetically highly activated breaking 2D bonds and remaking 3D chemical bonds, known as epitaxial stabilization[51], giving such heterovalent superlattices a higher chance to be made.

Based on first-principle calculations, we investigated the competition between stability and topological transition in lattice-matched heterovalent superlattices InSb/CdTe. We found that with increasing thickness of the InSb layer, the superlattices grown on [111] and [100] directions tend to have energy-lowering interfacial atomic exchanges, thus reducing the built-in polar field of abrupt interfaces. On the other hand, in [111] and [100] abrupt superlattices, as the InSb layer going thicker the built-in field could induce a NI-TI transition with a large excitation gap up to 156 meV as well as giant Rashba effect. Generally, accompanied with larger field is the cost of higher energy and larger possibility of reconstruction. Therefore, one can design heterovalent III-V/II-VI superlattices with certain sublayer thicknesses that have sufficient field to have a NI-TI transition but thermodynamically not too high in energy. The fact that such heterostructure TIs are composed of normal semiconductor or insulator building blocks that are not TIs in the individual bulk forms illustrates the potential of circumventing the need to discover TIs exclusively in high-Z compounds. Finally, our work illustrates how to make realistic predictions on TI by co-evaluating the competition between stability and property, and stimulate the investigation of novel functionality related to the topological nature of such recently made heterostructures with previously unmeasured properties.

**Methods**

*Total energy:* The calculations were performed by density functional theory (DFT) where the geometrical and total energies are calculated by the projector-augmented wave (PAW) pseudopotential [52] and the exchange correlation is described by the generalized gradient approximation of Perdew, Burke and Ernzerhof (PBE) [53] as implemented in the Vienna ab initio package (VASP) [54]. The plane wave energy cutoff is set to 450 eV, and the electronic energy minimization was performed with a tolerance of $10^{-5}$ eV. All the lattice parameters are fixed to the experimental value of InSb (6.47 Å), while the atomic positions were fully relaxed with a tolerance of 0.01 eV/Å.

*Electronic structure:* The PBE exchange correlation underestimates the band gap of both InSb and CdTe bulk, so for electronic structure calculation we choose the meta-GGA exange potential mBJ (modified Becke-Johnson) [55], which is reported to yields band gaps with an accuracy similar to hybrid functional (HSE) [56] or GW methods. The mBJ potential is a local approximation to an atomic exact-exchange potential plus a screening term, with their weight parameter CMBJ determined by the self-consistent electron density. For bulk InSb and CdTe the calculated parameter CMBJ are 1.21 and 1.24, respectively. The comparison of band gaps for bulk InSb and CdTe using different methods and with experimental value is shown in Table. S1. We found that the results of mBJ functional give good agreements with the experiments. Spin-orbit coupling is calculated self-consistently by a perturbation $\sum_{i,l,m} V_l^{SO} \vec{L} \cdot \vec{S} |l,m\rangle_{i\ i}\langle l,m|$ to the pseudopotential, where $|l,m\rangle_i$ is the angular momentum eigenstate of *i*th atomic site [57]. The atomic projection on band structure is calculated by projecting the wave functions with plan wave expansion on the orbital basis (spherical harmonics) of each atomic site.

*Topological invariant Z₂:* Here we use the method of the evolution of Wannier charge centers (WCCs) [40, 41] to calculate $Z_2$. The method is based on Wannier functions described as:

$$|Rn\rangle = \frac{i}{2\pi} \int_{-\pi}^{\pi} dk e^{ik(R-x)} |u_{nk}\rangle, \qquad (4)$$

which depends on a gauge choice for the Bloch states $|u_{nk}\rangle$. The WCC is defined as the mean value of the position operator $\bar{x}_n = \langle 0n|\hat{X}|0n\rangle$. For obtaining the WCCs, we follow the scheme proposed by Yu *et al* [40]. Fixing $k_y$, the maximally localized Wannier function can be obtained as eigenstates of position operator projected in the occupied subspace as follow:

$$\hat{X}_p(k_y) = \begin{bmatrix} 0 & F_{0,1} & 0 & 0 & \cdots & 0 \\ 0 & 0 & F_{1,2} & 0 & \cdots & 0 \\ 0 & 0 & 0 & F_{2,3} & \cdots & 0 \\ \vdots & \vdots & \vdots & \vdots & \vdots & \vdots \\ 0 & 0 & 0 & 0 & \cdots & F_{Nx-1,Nx-2} \\ F_{Nx-1,0} & 0 & 0 & 0 & \cdots & 0 \end{bmatrix}, \qquad (5)$$

with $F_{i,i+1}^{mn} = \langle m, k_{xi}, k_y | n, k_{xi+1}, k_y \rangle$. The eigenvalue of the projected position operator can be solved by the transfer matrix method:

$$D(k_y) = F_{0,1} F_{1,2} F_{2,3} \cdots F_{Nx-2,Nx-1} F_{Nx-1,0}, \qquad (6)$$

where the dimensionality of $D(k_y)$ is the number of occupied pairs. The eigenvalues of $D(k_y)$ are $\lambda_m^D(k_y) = e^{i\bar{x}_n(k_y)}$. For a dense mesh grid that fulfill, in the limit of an infinite lattice $\hat{X} \to i\frac{\partial}{\partial k_x}$ and $Z_2$ can be written as:

$$Z_2 = [\sum_\alpha \bar{x}_\alpha^I(TRIM_1) - \bar{x}_\alpha^{II}(TRIM_1)] - [\sum_\alpha \bar{x}_\alpha^I(TRIM_2) - \bar{x}_\alpha^{II}(TRIM_2)], \quad (7)$$

with $\alpha$ a band index of the occupied states, and superscripts I and II being the Kramer partners. Equation (7) explains that an odd number of switching between WCCs would make the $Z_2$ number odd, unveiling its topological nature.

**Acknowledgements**

We are grateful for the helpful discussions with Yong-Hang Zhang, David Smith and Robert Nemanich from Arizona State University on the growth of InSb/CdTe superlattice. Work of Q.L. and A.Z. on calculation of basic properties (including configuration study, total energy and band structure) was supported by Office of Science, Basic Energy Science, MSE division under grant DE-FG02-13ER46959 to CU Boulder. Work of Q.L., X.Z., L.B.A. and A.Z. on calculation of topological properties are supported by NSF Grant titled "Theory-Guided Experimental Search of Designed Topological Insulators and Band-Inverted Insulators" (No. DMREF-13-34170).

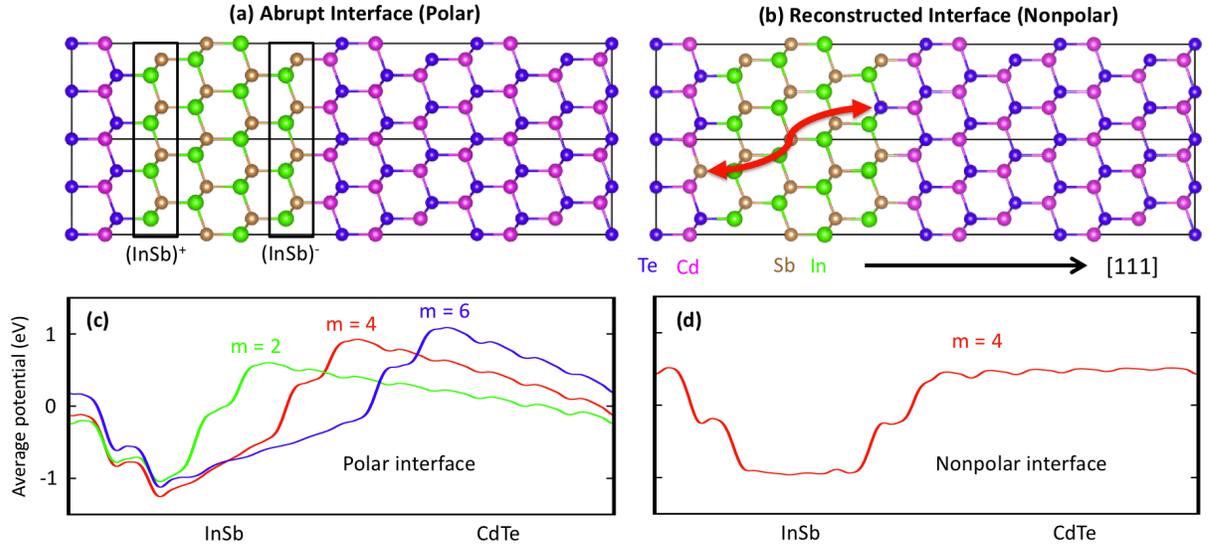

Figure 1: $(InSb)_m/(CdTe)_n$ [111] superlattices. Crystal structure of $(InSb)_4/(CdTe)_8$ demonstrating (a) abrupt polar interface and (b) reconstructed nonpolar interface with atomic exchange. The Sb-Te atom exchange is indicated by the red arrow. We denote the InSb layer at the donor interface and acceptor interface as $(InSb)^+$ and $(InSb)^-$, respectively. Average electrostatic potential of different configurations of $(InSb)_m/(CdTe)_n$ superlattices: (c) abrupt interface (polar) and (d) charge-compensated interface (nonpolar) by atomic exchange.

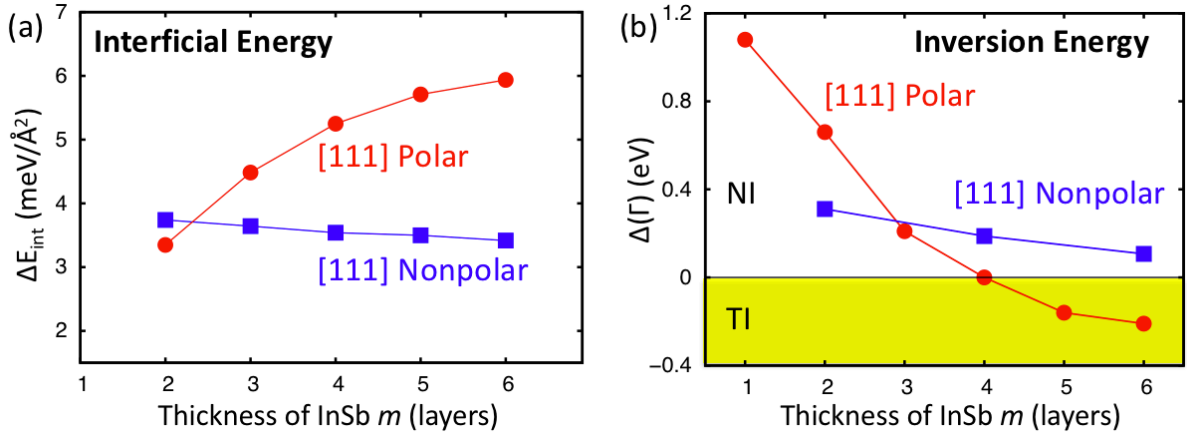

Figure 2: (a) Interfacial energy of abrupt polar (blue) and atomically exchanged nonpolar (red) configurations with respect to phase separation as a function of the thickness of InSb ML $m$. (b) Inversion energy $\Delta(\Gamma)$ between the conduction and valence bands as a function of the thickness of InSb ML $m$. Note that the atomically exchanged nonpolar configuration has no band inversion, so $\Delta(\Gamma)$ is also its excitation gap $E_g$.

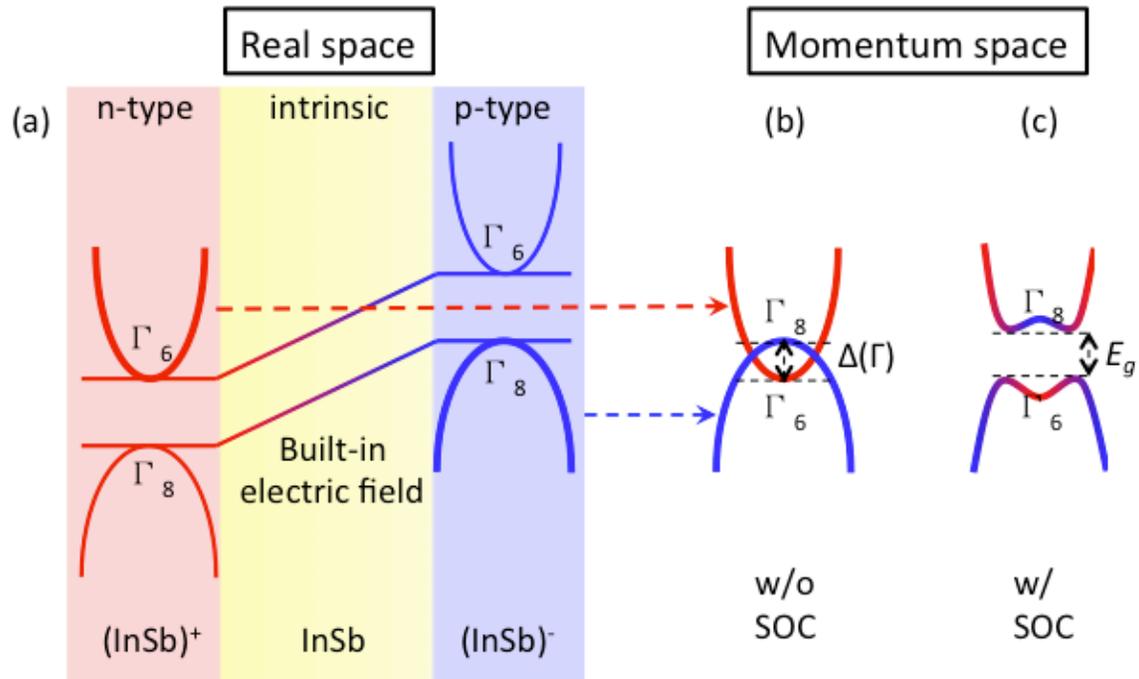

Figure 3: (a) Schematic band structure and real space potential alignment of *p-i-n* junction between the InSb layers with positively charged bonds and negatively charged bonds as $(InSb)^+$ and $(InSb)^-$, respectively. (b) The resulting k-space band structure with a band inversion at the Γ point (without SOC), and (c) the excitation gap opened by SOC. The bands with red and blue colors denote their origin from $(InSb)^+$ and $(InSb)^-$, respectively.

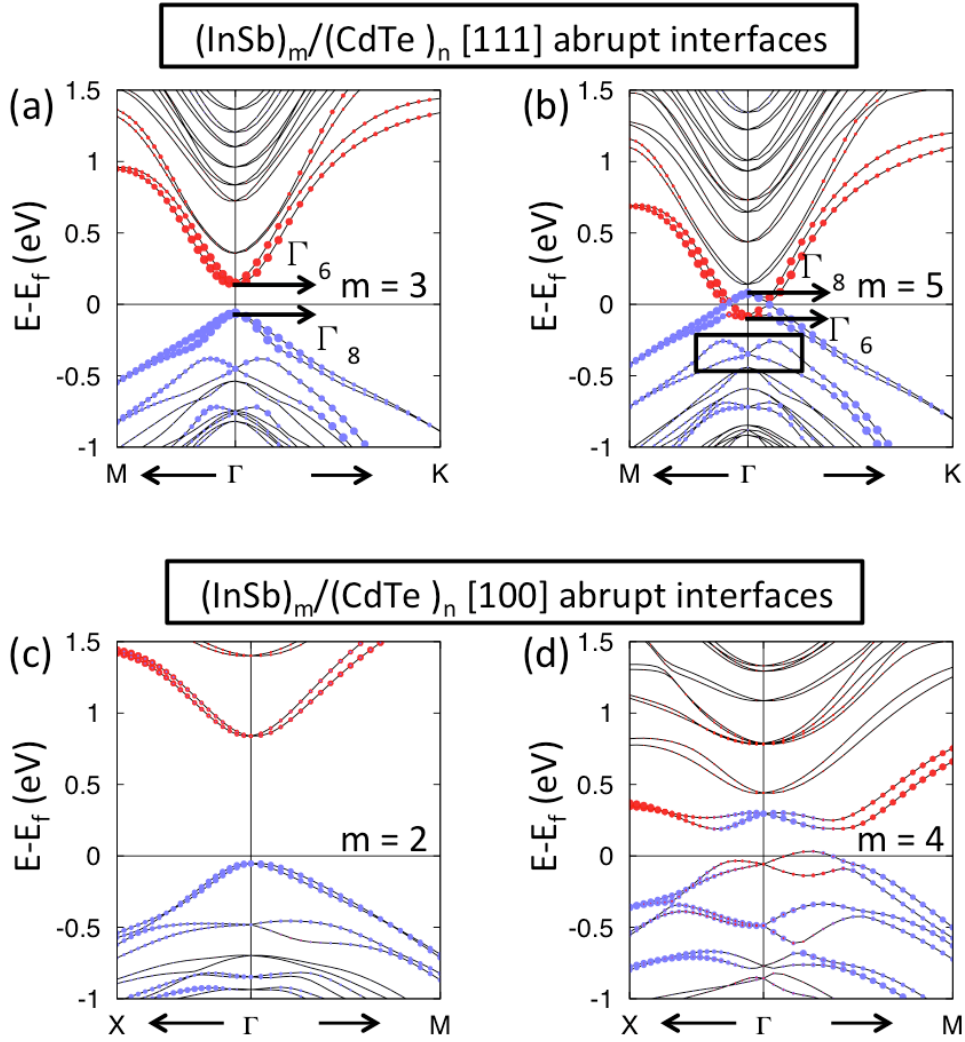

Figure 4: Band structure of [111] abrupt $(InSb)_m/(CdTe)_n$ superlattices with (a) $m = 3$ and (b) $m = 5$. The black frame in (b) indicates the subbands manifesting giant Rashba splitting. Band structure of the [100] abrupt $(InSb)_m/(CdTe)_n$ superlattices with (c) $m = 2$ and (d) $m = 4$. Red and blue dots denote atomic projection on to s state of $(InSb)^+$ and p state of $(InSb)^-$, respectively.

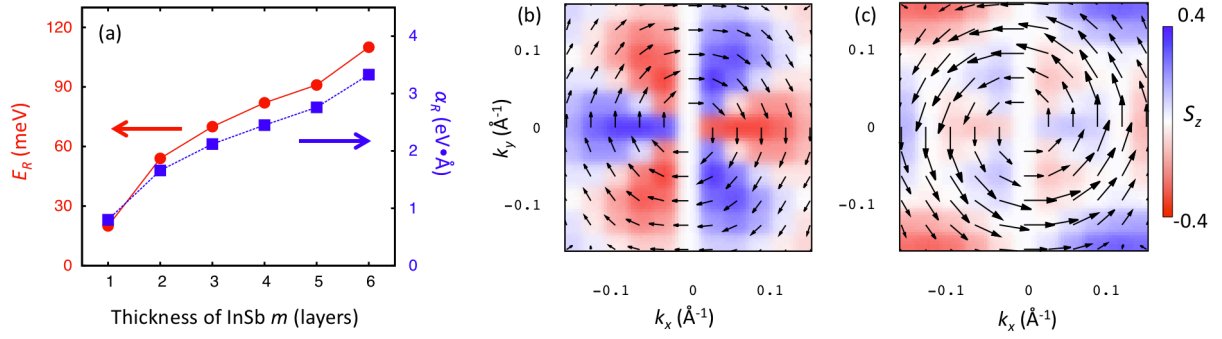

Figure 5: (a) Rashba energy $E_R$ and Rashba parameter $\alpha_R$ as a function of the thickness of InSb well $m$ in abrupt $(InSb)_m/(CdTe)_n$ superlattices. (b-c) Helical spin textures for (b) upper band and (c) lower band of the subbands below VBM in abrupt $(InSb)_5/(CdTe)_7$ superlattice, indicated by the black frame in Figure 3b. The background color indicates the out-of-plane spin component $S_z$.

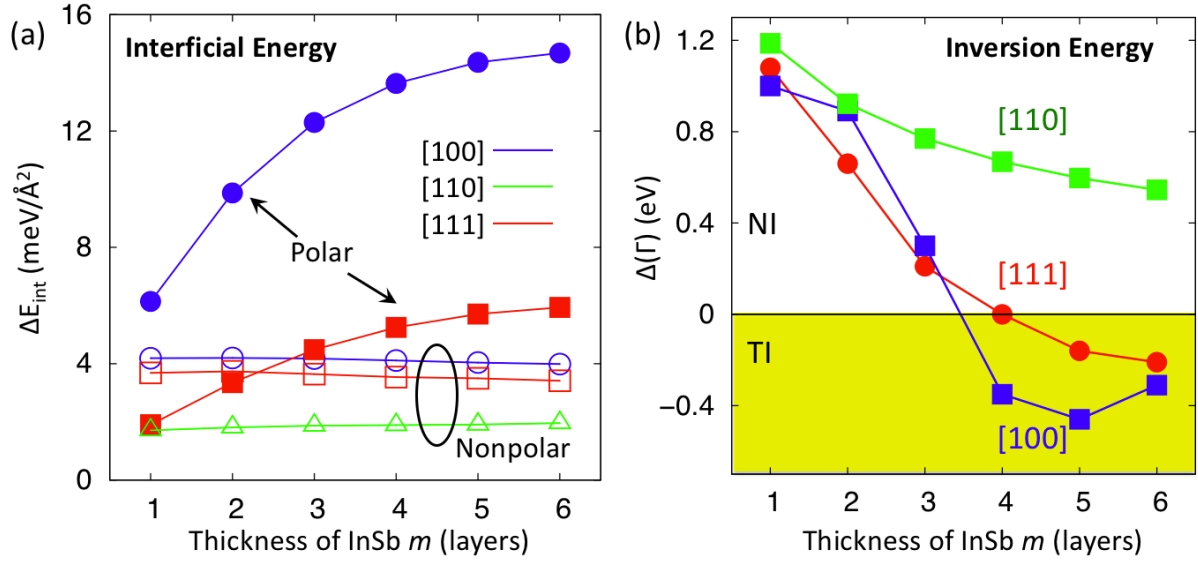

Figure 6: (a) Interfacial energy of InSb/CdTe superlattices for different configurations and directions. The solid and open symbols denote polar and nonpolar configurations, respectively. (b) Inversion energy $\Delta(\Gamma)$ of InSb/CdTe abrupt superlattices as a function of the thickness of InSb layers $m$. Note that for [110] direction, the abrupt superlattice is still nonpolar.